\documentclass[10pt]{article}
\usepackage[letterpaper]{geometry}
\usepackage{hicss51}
\usepackage{times}
\usepackage[none]{hyphenat}
\usepackage{url}
\usepackage{latexsym}
\usepackage{indentfirst}
\usepackage{graphicx}
\graphicspath{{images/}}

\usepackage[T1]{fontenc}
\usepackage[utf8]{inputenc}
\usepackage{algpseudocode}
\usepackage{algorithm}
\usepackage{listings}
\usepackage{color}
\usepackage{xspace}
\usepackage{url}
\usepackage{multicol}
\usepackage{booktabs}
\usepackage{graphicx}
\usepackage{hyperref}
\usepackage{enumerate}
\usepackage{inconsolata}
\usepackage{microtype}

\usepackage{todonotes}
\usepackage{silence}
\usepackage{subcaption}
\usepackage{graphicx}
\usepackage{hyperref}

\newcommand{\solntxt}{SFC\-tree\xspace}
\newcommand{\solutn}{\texttt{\solntxt}\xspace}

\newcommand{\problemtxt}{SFC design problem\xspace}
\newcommand{\probm}{\texttt{\problemtxt}\xspace}

\renewcommand{\paragraph}[1]{\hfill\break\noindent\textbf{#1}\xspace}

\colorlet{punct}{red!60!black}
\definecolor{delim}{RGB}{20,105,176}
\definecolor{keyword}{RGB}{48,0,211}
\colorlet{numb}{magenta!60!black}

\newcommand{\kwd}[1]{{\color{keyword}\textbf{#1}}}
\newcommand{\hid}[1]{{\color{darkgray}#1}}

\lstdefinelanguage{json}{
    basicstyle=\footnotesize\ttfamily,
    mathescape=true
    stepnumber=1,
    showstringspaces=false,
    keywordstyle = {\bfseries\color{keyword}},
    keywords = {src,dst,node_value,qos_type,qos_thr,qos_value,qos,vnfList,%
    dupList,prox_to_src,prox_to_dst},
    keywordstyle = [2]{\bfseries\color{punct}},
    morekeywords = [2]{string,array,enum,integer,boolean,items,int},
    keywordstyle = [3]{\color{darkgray}},
    morekeywords = [3]{type,object,properties,item,description},
    literate=
     *{0}{{{\color{numb}0}}}{1}
      {1}{{{\color{numb}1}}}{1}
      {2}{{{\color{numb}2}}}{1}
      {3}{{{\color{numb}3}}}{1}
      {4}{{{\color{numb}4}}}{1}
      {5}{{{\color{numb}5}}}{1}
      {6}{{{\color{numb}6}}}{1}
      {7}{{{\color{numb}7}}}{1}
      {8}{{{\color{numb}8}}}{1}
      {9}{{{\color{numb}9}}}{1}
      {:}{{{\color{gray}{:}}}}{1}
      {"}{{{\color{gray}{"}}}}{1}
      {,}{{{\color{punct}{\textbf{,}}}}}{1}
      {\{}{{{\color{gray}{\textbf{ \{ }}}}}{1}
      {\}}{{{\color{gray}{\textbf{ \} }}}}}{1}
      {[}{{{\color{delim}{\textbf{[}}}}}{1}
      {]}{{{\color{delim}{\textbf{]}}}}}{1},
}

\newtheorem{definition}{Definition}
\newtheorem{theorem}{Theorem}
\newtheorem{example}{Example}

\title{Constraint programming for flexible Service Function Chaining deployment
}

\begin{document}


\author{
Tong Liu$^{\dagger,\star}$,
Franco Callegati$^{\dagger}$,
Walter Cerroni$^{\dagger}$,
Chiara Contoli$^{\dagger}$,
Maurizio Gabbrielli$^{\dagger,\star}$,
Saverio Giallorenzo$^{\diamond}$
\\ 
$^{\dagger}$Universit\`a di Bologna,
$^{\star}$INRIA,
$^{\diamond}$University of Southern Denmark
\\
\{\underline{\vphantom{g}t.liu},
\underline{franco.callegati},
\underline{\vphantom{g}walter.cerroni},
\underline{\vphantom{g}chiara.contoli}
\underline{maurizio.gabbrielli}
\}
\underline{@unibo.it},
\underline{saverio@imada.sdu.dk}
}



\maketitle

\begin{abstract}

Network Function Virtualization (NFV) and Software Defined Networking (SDN) are
technologies that recently acquired a great momentum thanks to their promise of
being a flexible and cost-effective solution for replacing hardware-based,
vendor-dependent network middleboxes with software appliances running on general
purpose hardware in the cloud. Delivering end-to-end networking services across
multiple NFV/SDN network domains by implementing the so-called Service Function
Chain (SFC) i.e., a sequence of Virtual Network Functions (VNF) that
composes the service, is a challenging task.

In this paper we address two crucial sub-problems of this task: i) the
language to formalize the request of a given SFC to the network and ii) the
solution  of the SFC design problem, once the request is received.  As for i)
in our solution the request is built upon the intent-based approach, with a
syntax that focuses on asking the user "what" she needs and not "how" it
should be implemented, in a simple and high level language. Concerning ii) we
define a formal model describing network architectures and VNF properties that
is then used to solve the SFC design problem by means of Constraint
Programming (CP), a programming paradigm which is often used in Artificial
Intelligence applications. We argue that CP can be effectively used  to
address this kind of problems because it provides very expressive and flexible
modeling languages which come with powerful solvers, thus providing efficient
and scalable performance. We substantiate this claim by validating our tool on
some typical and non trivial SFC design problems.

\end{abstract}

\section{Introduction}

Following the recent innovations brought about by Cloud Computing and resource
virtualization, current advances in communication infrastructures show an
unprecedented central role of software-based solutions~\cite{callegati2013sdn}.
On the one hand, Network Function Virtualization (NFV)~\cite{CommSurvNFV2016}
supports the deployment of network functions---e.g., load balancers, firewalls,
intrusion detection devices, and traffic accelerators---as pieces of software
running on off-the-shelf hardware. On the other hand, Software Defined
Networking (SDN)~\cite{CommSurvSDN2014} decouples the software-based control and
management plane from the hardware-based forwarding plane, turning traditional
infrastructures into fully programmable communication platforms. A SDN is hence
a network whose topology can be orchestrated dynamically. By taking advantage of the complementary features of NFV and
SDN it fosters the provision of flexible and cost-effective network services---from
now on, referred simply as \emph{services}.

As detailed in Section~\ref{sec:background}, in an NFV/SDN framework, services
are deployed as Service Function Chains (SFC)~\cite{RFC-SFC}, i.e., the
concatenation of some basic functions, typically running in some form of
virtual environment (virtual machine, container etc.). These are called
Virtual Network Functions in short VNFs. Essentially, an SFC corresponds to
the sequence of VNFs that a traffic flow traverses from its source to its
destination. In this context, multiple network configurations can coexist over
the same physical infrastructure, bypassing the need for specialized hardware
and physical network reconfigurations. Moreover the software-based SFCs can be
instantiated, controlled, modified, and removed over a small time scale which
is impossible to achieve in traditional networks typically requiring physical
or manual reconfiguration to modify topology and/or forwarding. However, one
of the main problems linked to SFC planning is that it is complex to define
and apply SFC configurations that both respect multiple domain-level
properties (QoS, etc.) and also avoid misbehaviors over contrasting or
incompatible service desiderata. This calls for both suitable, high-level
languages to easily describe SFC requests and for tools to efficiently design
SFC---once the request is received---given  the available VNFs and network
resources.

\paragraph{Contribution.} Answering this call, in this paper we propose two
contributions. The first is a model to describe both SFC user requests and the
holding domain-level constraints over a multi-domain network scenario---since
the model is intended for (possibly automated) user interaction (both
customers and network administrators) it is expressed using the familiar
JavaScript Object Notation (JSON).  The second is a tool based on Constraint
Programming (CP) which solves the SFC design problem. The tool uses a MiniZinc
specification which is a direct  translation of the  JSON  specification.
While there exists another paper~\cite{layeghy2016scor} using CP techniques
for routing problems, ours is the first proposal of applying CP to the SFC
design problem in its full generality. We argue that CP can be effectively
used to address this kind of problems, as it provides very expressive and
flexible modeling languages to harness the complexity of SFC design. This,
together with the outstanding performance of modern CP solvers, has promising
aspects in terms of scalability, opening the market to operators offering
ad-hoc just-in-time SFC configurations to users. To substantiate our claims we
validated our tool by  solving some typical and non-trivial SFC design
problems and considering its performance.

In the remainder of the paper, in Section~\ref{sec:background} we provide background
knowledge and a detailed description of NFV/SDN-based frameworks, introducing
the elements of the problem. In Section~\ref{sec:problem_def} we set the general
problem framework and present our model to specify user desiderata and
domain-level properties. In Section~\ref{sec:problem_modeling} we describe how to
translate a given model into a MiniZinc finite domain specification, reporting
in Section~\ref{sec:experiments} validation experiments and performance results.
Finally, in Section~\ref{sec:related} we consider related work, we draw conclusions,
and delineate future work.

\section{Application Context: NFV/SDN Networking}
\label{sec:background}
In this section we introduce our application context and its
elements (identified by the paragraph titles).

\begin{figure}[tb]
\centering
\includegraphics[width=0.45\textwidth]{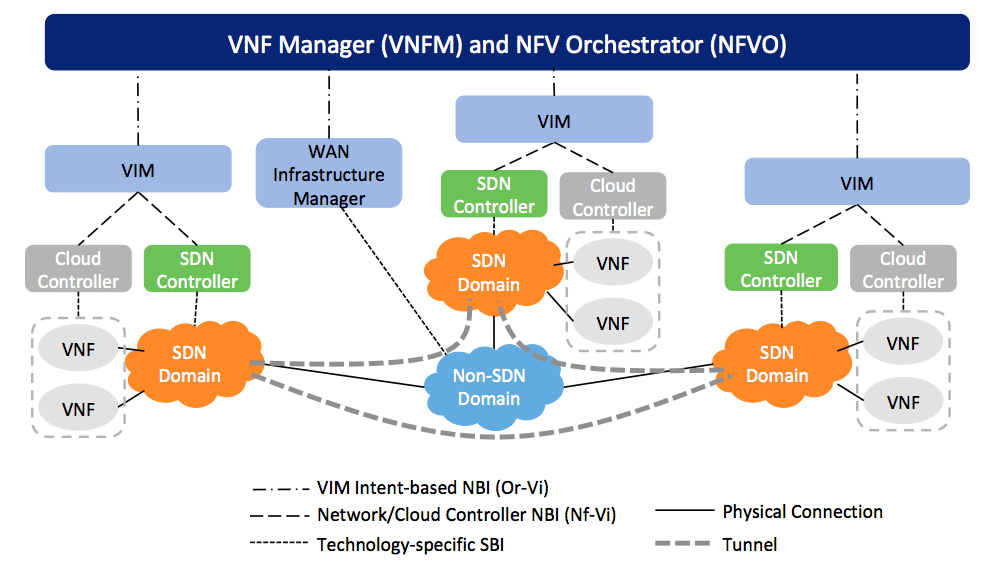} 
\caption{General concept of MANO.}
\label{fig:MANO}
\end{figure}

NFV/SDN paradigms promise to revolutionize network management through the
concept of \emph{network programmability}, i.e., the possibility to run network
services in a similar way as running software in a computer. Indeed, traditional
network functions are bound to hardware devices, in which actions like
instantiating a new service or modifying a service instance are rather complex
and require specialized operations. Contrarily, the combination of recent NFV/SDN technologies paves the way to fully programmable communication networks.
The expected benefits of programmable networks are reduced operation costs, as well
as increased flexibility and responsiveness.
\paragraph{Network Function Virtualization.}
In NFV network functionalities, mostly implemented by means of
dedicated appliances (\emph{middleboxes}, like firewalls, NATs,
packet inspectors, traffic conditioners, etc.) are turned into software
applications, called Virtual Network Function (VNF). These are shipped inside
virtual machines or containers and hosted into cloud computing infrastructures
equipped with off-the-shelf hardware (i.e., not specialized for a specific
networking function)~\cite{CommSurvNFV2016}.
%
%
%
\paragraph{Software Defined Networking.}
SDN decouples the network control plane from the data forwarding plane. The
former is placed into a so called \emph{SDN controller}, defining all the
forwarding logics in a centralized way and injecting them into the networking
devices. The main protocol proposed for SDN is Openflow~\cite{OpenFlow}, which
is designed to support the dialog between network controllers and appliances.
\paragraph{The ETSI NFV-MANO Framework.}
NFV 
became subject of standardization by ETSI in the NFV Management and
Orchestration (MANO) framework. ETSI launched the initiative by bringing
together seven leading telecom operators in 2012. Currently over 300 individual
companies~\cite{ETSI-companies}, including many global service providers, joined
the initiative, which is the reference standardization framework in this field.
We provide in Fig.~\ref{fig:MANO} a conceptual representation of the approach
proposed by the ETSI NFV-MANO framework---from now on called
MANO~\cite{ETSI-MANO}. In MANO, VNFs are deployed over a set of cloud data centers
that may be either closely or remotely located, depending on the specific
service implementation scenario. The data centers are managed by a specific cloud infrastructure management system chosen by the owner/provider, e.g., the
renowned OpenStack \cite{OpenStack} platform, while general networking services
are managed by SDN controllers. MANO addresses both cloud and network controllers as
Virtualized Infrastructure Managers (VIMs).
\paragraph{The NorthBound Interface.}
The components in Fig.~\ref{fig:MANO} must interact by means of suitable
Application Programming Interfaces (APIs) and, roughly speaking, the API
offered by a given functional block to the one that is logically above it
(providing increased abstraction) is usually called a \emph{NorthBound
Interface (NBI)} while the interface with one logically below (closer to the
specific implementation) is called a \emph{SouthBound Interface
(SBI)}\footnote{For completeness, interfaces between functional blocks at the
same architectural level are usually addressed as East/West-bound
interfaces.}.
\paragraph{The Service Function Chain.}
In this context, a \emph{service} is a specific combination of VNFs and
communication capabilities that are requested by a user and that must be
implemented in the available infrastructure.\footnote{Here, \emph{users} may
either be \emph{customers} (residential or business) requiring a specific
networking service or \emph{network operators} configuring specific
services for their customers.}
This is the \textit{Service Function Chain (SFC)}, i.e. the implementation of a
composite service as the concatenation of basic services, typically implemented
via VNFs. For instance an SFC could be the sequence of a NAT and a Firewall at
the edge of the provider network, serving a set of customers. In essence, an SFC
is the series of VNFs that a traffic flow must traverse from its
source to its destination. \emph{Thanks to the capabilities offered by SDN and
NFV, SFCs can be dynamically controlled and modified over a relatively small
time scale, both increasing the flexibility of service provisioning and reducing
the management burden.}
\paragraph{SFC deployment planning.}
The aspect we focus in this paper is SFC deployment planning, also called
Service Function Chaining (SF-Chaining).
Within a single technological and administrative domain, e.g., a single data
center, SF-Chaining can be successfully achieved with the help of the native
domain management system, i.e. the VIM~\cite{callegati2016sdn}.
\begin{figure}
\centering
\includegraphics[width=0.4\textwidth]{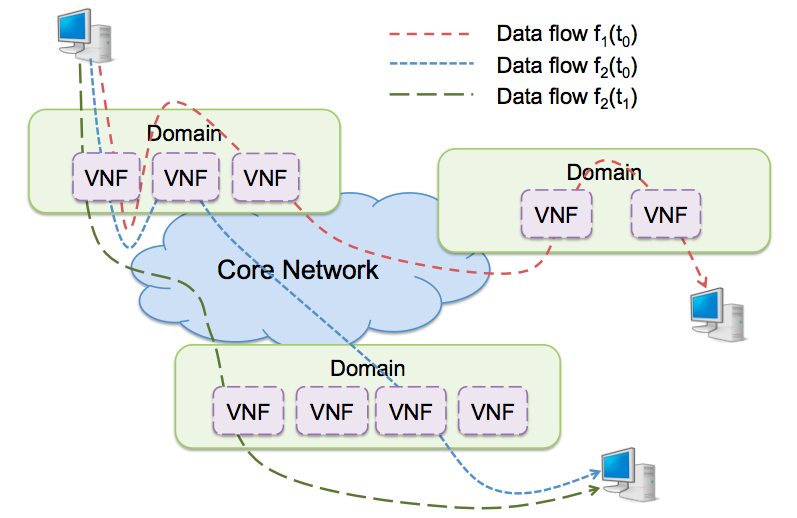} 
\caption{General example of dynamic Service Function Chaining.}
\label{fig:SFC}
\end{figure}
However, when the SFC spans across multiple network domains, (c.f.,
Fig.~\ref{fig:SFC}) each owned by a different player and characterized by
different technology stacks, the dimensional
and logical complexity of the problem increases. With many domains and many VNFs
per domain the space of possible solutions to a specific SF-Chaining problem
becomes very large as formally shown in
the following section. Moreover the specification of the
SF-Chaining request in a general way, that can be mapped over the various
domains is also non
trivial ~\cite{NetSysMD2NFV2015,NOMSDISCO2014,SIGCOMMUNIFY2015}.

MANO provides a general architectural framework for the implementation of NFV
but does not provide implementation details for the various interfaces of
logical levels, that are still matter of study and testing. 

Regarding the specification of the
SF-Chaining request, solutions have been recently proposed to implement a
vendor-agnostic, and interoperable NBI interface for the MANO according to the
\emph{intent-based} approach \cite{cohen2013intent}. Very briefly the intent-based
approach goal is to provide a semantic at the interface that allows the user to focus on \emph{what} he/she wants to achieve and not
on \emph{how} it will be implemented, thus hiding all the technology-specific
details and making the service request as general as possible. In this work we
extend and better formalize this approach by providing a general schema for the
semantics of the interface that can be easily translated into technology
dependent specifications.

While the intent-based specification solves the problem of applying a global
plan over multiple domains, it does not answer the problem of engineering the
SF-Chaining, which instead need to consider: \emph{SFC design}, i.e.,
selecting the set of VNFs to be chained to implement the SFC, with
the goal of optimizing some notion of cost; \emph{VNF activation and
placement}, i.e., where to execute VNFs when more options
are available, for instance with the goal to maximize performance or
distribute the workload.

SF-Chaining is a crucial part of the Resource Allocation problem in an NFV environment and has been mostly studied by means of Mixed Integer Linear Programming \cite {HB16}.  Unfortunately the complexity of the problem makes such solutions viable just for small networks. Usually heuristics are proposed and tailored to some specific optimization goal, thus limiting their applicability or generality.
The problem is that, when designing an SFC, beside standard shortest-path
problems, one has to solve additional constraints arising from the specific
nature of the service functions involved. For example, if a Virtual Private
Network (VPN) function is present, which encrypts a message before it leaves the
source domain, then a complementary VPN function should appear before the final
destination, to decrypt the message.

In this work we propose an efficient, general and scalable tool, based on
Constraint Programming (CP), for the engineering of SFC plans over multiple
domains. We will show that complex SFC plans can be computed in a small time-frame, turning the
engineering and application of SFC plans from a manual, time-consuming activity
to an automatic and just-in-time task.

\section{Problem Definition}
\label{sec:problem_def}

With reference to what explained above, in this section we set the general
problem framework following the schematic presented in Fig.~\ref{fig:SFC}. In
particular we assume the following.

\begin{itemize}
\setlength{\itemsep}{-4pt}
\item \emph{Network architecture}. The network is divided into a number of
\emph{Domains}, defined according to administrative and/or technological
boundaries. For the purpose of this work a Domain is an infrastructure that is
managed homogeneously by a single actor. The Domain has one or a set of Virtual
Infrastructure Managers that are properly coordinated and thus acts as a single
entity. The resources of the Domain are managed as a whole.

\item \emph{Inter-Domains interconnection}. We assume that the various Domains
are interconnected by Domain border gateways and interconnection links. Domain
interconnections may be at the geographical as well as at the local level,
depending on topological and administrative constraints. Domain interconnection
can be related to some form of QoS objective, either cost, latency, bandwidth
availability, etc. depending on the specific scenario.

\item \emph{Intra-Domains interconnection}. The networking among VNFs of the
same domain is not a subject of this work. We assume that, within a domain,
connectivity is granted at a level of Quality of Service sufficient for the
purpose. If the various domains are data centers, their management platforms
provision the resources needed in terms of computation, networking etc.

\item \emph{VNFs}. The Virtual Network Functions are devoted to specific
networking tasks. In this work we assume that one VNF performs just one task,
therefore we will talk of VNF types to specify which tasks are performed. The VNF
types considered in the following are briefly described below.

\item \emph{VNF location}. VNFs are executed in the data centers hosted in the
various Domains. In principle the Domains are not homogeneous in terms of
connectivity, computing capabilities and functionalities, therefore a Domain may
or may not be suitable to execute some VNFs. Moreover it may be that a given VNF
has to be executed into a specific domain. Without loss of generality, we
restrict the choice of the location of each VNF in an SFC to three options: the
source Domain, the destination Domain or unspecified; the latter meaning that
the VNF can be located in any available Domain, including source and
destination.
\end{itemize}

The set VNF types is a set of network functions that we consider to be part of
common networking practice, obviously the work can be extended to include other
types of VNFs.

\begin{itemize}
\setlength{\itemsep}{-4pt}

\item \emph{Deep Packet Inspector (DPI)}. Looks into the content of the packets
and takes specific forwarding decisions according to specific predefined
patters.

\item \emph{Network Address Translator (NAT)}. Translates IP addresses mostly
used to interconnect ares with private IP addressing from the public Internet.

\item \emph{Traffic Shaper (TS)}.  May enforce specific packet and/or bit rate limitations to a traffic flow.

\item \emph{Wide Area Network Accelerator (WANA)}. Compresses packet content to
provide higher transfer speed.

\item \emph{Virtual Private Network Endpoint (VPN)}. Encrypts data flows and
authenticate users over a specific public network section.
\end{itemize}
Note that {\em gateway} VNFs do not appear in the user desiderata, however,
since they provide inter-domain connections, we will also consider them among
VNFs.

\subsection{Service Function Chain specification}


In the remainder, to distinguish between customer and network operator SFC
desiderata, we call the former \emph{user requests} and the latter \emph{domain
constraints}. In order to provide a concrete and simple model for specifying SFC user
requests, immediately usable in practice, we rely on the
JSON~\cite{crockford2006} notation, defining the model using the generic
formalism of JSON Schema~\cite{json_schema} as follows.

\begin{definition}[SFC user request]\label{def:userreq} A Service Function Chain
user request is any JSON specification compliant with the JSON Schema below
(indented and grayed-out to ease reading), where we assume that the
cardinalities of \lstinline{vnfList}, \lstinline{prox_to_src}, and
\lstinline{prox_to_dst} are equal.
\begin{lstlisting}[language=json,mathescape=false,escapeinside={<}{>}]
{"<\kwd{VNFs}>":{"type":"array","items":{"type":"string",
 "enum":["DPI","NAT","TS","WANA","VPN"]}},
 "<\kwd{Mask}>":{"type":"array","item":{"type":"boolean"}},
   "type":"object","properties":{
  "src":{"type":"string"},
  "dst":{"type":"string"},
  "qos":{"type":"string"},
  "qos_type":{"type":"string"},
  "qos_thr":{"type":"string"},
  "qos_value":{"type":"integer"},
  "vnfList":{"<\hid{\$ref}>":"<\hid{\#/}\kwd{VNFs}>"},
  "dupList":{"<\hid{\$ref}>":"<\hid{\#/}\kwd{VNFs}>"},
  "prox_to_src":{"<\hid{\$ref}>":"<\hid{\#/}\kwd{Mask}>"},
  "prox_to_dst":{"<\hid{\$ref}>":"<\hid{\#/}\kwd{Mask}>"}}}
\end{lstlisting}

\end{definition}

Briefly, the highlighted elements in Definition~\ref{def:userreq} represent:
\begin{itemize}
\setlength{\itemsep}{-4pt}
\item \lstinline|src| and \lstinline|dst| the start and target domain of the
service chain;
\item \lstinline|qos| the QoS feature to be provided with the service chain;
\item \lstinline|qos_type| a high-level unique identifier of a QoS metric;
\item \lstinline|qos_thr| the QoS threshold to be applied to the specified
metric;
\item \lstinline|qos_value| the value assigned to the threshold;
\item \lstinline|vnfList| is the ordered list of VNFs to be traversed for the
requested service. We \lstinline{enum}erate them in type \kwd{\lstinline{VNFs}}
as strings representing the VNFs we support in our model (and mentioned at the
beginning of Section~\ref{sec:problem_def});
\item \lstinline|dupList| is the set of VNF types where the traffic needs to be
duplicated.
\end{itemize}

Finally, \lstinline|prox_to_src| and \lstinline|prox_to_dst| are
\kwd{\lstinline{Mask}}s on the \lstinline|vnfList|, i.e., they are arrays of
booleans with the same cardinality of \lstinline|vnfList| that indicate if a VNF
should be respectively located in the domain of the \lstinline|src| or of the
\lstinline|dst|.

\begin{example}\label{ex:user_desiderata} To complete
Definition~\ref{def:userreq}, we report an example of SFC user request. In the
code below, the user requests a chain between domains
\lstinline{s} and \lstinline{d}, indicating a \lstinline{qos} on the
\lstinline{speed} of the connection, measured in terms of \lstinline{bandwidth}
with a threshold of \lstinline{90}\% on the \lstinline{throughput} of
transmitted data. The service request consists of (in this order): a
\lstinline{DPI} (whose traffic is duplicated, as per \lstinline{dupList}), a
\lstinline{VPN} in the domain of \lstinline{s} and a complementary
\lstinline{VPN} function in the domain of \lstinline{d}.

\begin{lstlisting}[language=json]
{"src":"s","dst":"d","qos":"speed",
"qos_type":"bandwidth",
"qos_thr":"throughput","qos_value":90,
"vnfList":["DPI","VPN","VPN"],
"dupList":["DPI"],"prox_to_src":[1,1,0],
"prox_to_dst":[0,0,1]}
\end{lstlisting}
\end{example}

In the next section, we explain how we combine the parameters above are to
define the solution to an SFC planning problem.

\subsection{SFC design problem}
\label{sub:SFCdesign}

In order to formalize the \probm we represent a network architecture in abstract
terms as a directed graph $G(V,L)$ with a set $V$ of labeled nodes, ranged over
by $v_1,v_2 \dots$, which represent the VNFs and a set $L = \{(u,v)|\forall u,v
\in V \wedge u \neq v \}$ of labeled edges---in the remainder called 
\emph{arcs}---ranged over by $l_1, l_2, \dots$, which represent links among
different VNFs. The level of a node $v$ denote the type of functionality
provided by the specific VNF $v$ in set $T$, ranged over by $t_1, t_2, \dots$,
and we assume that there exists a total function $Type: V\rightarrow T$ which,
for any VNF $v \in {V}$, returns its label (i.e., its type). We distinguish
between a VNF and its type because different VNFs, also in the same domain,
can offer the same functionality and have the same type. Nevertheless, when no
ambiguity arises, we will  identify a VNF with its type. For example, in the
service chain request provided by the user, the list of VNF which is provided
is, strictly speaking, the list of VNF types which are required (the user is
interested in a functionality, not in the specific component implementing it).
Label of arcs denote costs of the arcs and we indicate by $c_{u,v}$ the cost
of an arc $(u,v).$  Paths are defined as usual\footnote{For the notions on
graphs not directly  defined here please see
\cite{cormen2009introduction,deo2017graph}.}.

As we have seen in previous section, conceptually VNFs are organized in domains
that is, our graph is divided into several  sub-graphs. We represent this
structure by introducing a set $D$ of domains, ranged over by $d_1, d_2,
\ldots$, and assuming that there exists a total function Domain: $V\rightarrow
D$ which  for any VNF $v \in {V}$ provides its domain Domain($v$).
%
%
We assume that  each domain in our network has exactly one VNF providing the
(domain border) \emph{gateway} functionality.
%
%
In order to model the domain interconnection described above, we assume that
the set of arcs in our network consists of two types of arcs: those connecting
the gateway to all the other VNFs in the same domain (with cost 0) and those
connecting a gateway to all the gateways VNF appearing in the other domains,
with a positive cost.
%
%
We are now ready to define the notion of \solutn. Intuitively this represents
the chain of functions which, in a given network, satisfy the service request
expressed by the user. Note that we consider a tree rather then a simple path
because in some cases the chain of functions, beside a source and a target,
include some other terminating nodes which provide specific functionalities: for
example, a $\texttt{DPI}$ VNF has the task of logging messages and does not
participate in message routing. Moreover, nodes (VNFs) in the same domain are
represented as sons of a gateway.
%

\begin{definition}[\solutn] Given a directed graph ${G}( {V}, {L})$ representing
a network architecture, an \emph{\solutn}\footnote{The definition is parametric
w.r.t. the given graph, however we do not represent such a parameter explicitly,
to simplify the notation.} is a rooted tree $Tr$ which is  a subgraph of ${G}(
{V}, {L})$ and such that
%
%
the leafs of  $Tr$ are (labeled by) {\em VNFs} types different from {\em
gateway}, while the nodes that are not leafs are (labeled by) {\em gateway}.

%
\end{definition}

As a first approximation, our configuration problem consists in finding an
\solutn which satisfies the service request specified by the user in terms of
intents. There are however some additional, domain level, constraints on the
VNFs to be used in the SFC which are needed to obtain a correct solution. For
example, we may need to know whether a VNF $v$ needs to be "mirrored" , meaning
that when $v$  appears in a chain then another, dual,  VNF is needed in the same
chain (for example an encryption function needs later a decryption). Also,  some
quantitative information are needed at domain level, such as  lower and upper
bounds on the number of VNFs  of the same type in a given domain. These
additional constraints are not expressed by the intents of the users (who might
ignore the detailed domain structure of the network) but are introduced in a
middle layer before formulating the actual service request. As we have done for
SFC user request, we represent these constraints following the JSON Schema.

\begin{definition}[Domain-constraints] \label{definition:VNFcon} 
A Domain-constraint is a JSON specification compliant with the following JSON
Schema
\begin{lstlisting}[language=json,escapeinside=\<\>]
{"type":"array","items":{"type":"array",
"maxItems":4, "items":[{"type":"string",
"description":"a domain name"},
{"type":"string",
"enum":["DPI","NAT","TS","WANA","VPN"]},
{"type":"integer",
"description":"VFN type minimum quantity"},
{"type":"integer",
"description":"VFN type maximum quantity"}]}}
\end{lstlisting}
\end{definition}
In the JSON Schema above, we use the
\lstinline{"description"} attribute to hint the content of each element. A
Domain-constraint then represents a set of tuples $(d,t,m,n)$ where $d$ is a domain, $t$ is a VNF type, and $m,n$ are natural numbers, with the meaning that in
the domain $d$ there are at least $m$ and at most $n$ VNFs $v\in V$ having the
type $t$.
%

\begin{example}
To complete Definition~\ref{definition:VNFcon}, we report an example of a
Domain-constraint which could be imposed by domain administrators. Here
\lstinline{s} and \lstinline{d} are the source and destination domains of
Example~\ref{ex:user_desiderata} and we see that the administrator set to
\lstinline{1} and  \lstinline{2} the minimal a maximal number of
\lstinline{WANA} functions allowed in \lstinline{s}; the constraint specifies
also that a single \lstinline{DPI} function is required in  \lstinline{s} (i.e.,
minimal and maximal capacities coincide) and a single \lstinline{VPN} (and
\lstinline{NAT}) is required in the destination \lstinline{d}.
\begin{lstlisting}[language=json]
[["s","WANA",1,2],["s","VPN",5,10],
["s","DPI",1,1],$...$ ["other_dom",DPI,1,2], 
["other_dom","VPN",1,10], $...$ 
["d","VPN",1,1],["d","NAT",1,1]]
\end{lstlisting} 
  
\end{example}

Before defining formally our \probm we now need to define when an \solutn---that
intuitively represents a solution---satisfies the user request and the Domain
constraints. To this aim, we first provide the following definition.

\begin{definition} 
Assume that $R$ is an SFC user request specified as in Definition~
\ref{def:userreq} which defines the $\texttt{vnfList} = \{t_1,\dots,t_n\}$ and
a $\texttt{dupList} = \{e_1,\dots,e_m\}$. Then we define
$\texttt{request-tree(R)}$ as the tree $T(V,L)$ where the set of nodes is $V=
\{v_1,\dots,v_n\}$  with $Type(v_i) = t_i$, $\forall i \in [1,n]$ and the set of
arcs is $L= \{(v_i,v_j) | v_i,v_j \in V \wedge i < j \wedge Type(v_i) \notin
\texttt{dupList} \wedge ( ... nexists k, i<k<j, v_k \notin \texttt{dupList} )\}$.
\end{definition} 

Intuitively, given a user request $R$, $\texttt{request-tree(R)}$ is the tree
that represents the traversal order of the various VNFs, from the source to the
destination domain, to obtain a solution. We have a tree rather then a
sequence of VNFs because we take into account also the information provided by
$\texttt{dupList}$ which, as mentioned before, specifies when the traffic needs
to be duplicated before entering in a node (VNF).

\begin{example}
Given a user request which specifies $\texttt{vnfList} = \{a,b,c,d\}$ and
$\texttt{dupList} = \{b\}$, with $a$ in the source domain and $d$ in the
destination domain,  a $\texttt{request-tree}$ $T(V,L)$  consists of
$V=\{a,b,c,d\}$, $L=\{(a,b),(a,c),(c,d)\}$.\end{example}


Next we define the satisfaction of user request and domain constraints. In the
following we use the terminology and notation introduced in
Definitions~\ref{def:userreq} and~\ref{definition:VNFcon}. We also assume that
the last VNF specified in the user $\texttt{vnfList}$ is present in the
destination domain (if this were not the case we could introduce and additional
$Endpoint$ VNF but we prefer to avoid this in order to simplify the notation).

\begin{definition} 
We say that an \solutn $Tr(V_r,L_r)$ satisfies user request $R$ and domain
constraints $C$ if the following holds, where $\texttt{request-tree(R)} =
T(V,L)$ and $d_{src}, d_{dst}$ are the domains values specified in
$\texttt{dst}$ and $\texttt{src}$ of request $R$:
%
\begin{enumerate}[i)]
\setlength{\itemsep}{-4pt}
\item the domain of the root of $Tr$ is $d_{src}$
and there exists a leaf in $Tr$ whose domain is $d_{dst}$.
\item $V_r$ is the set $V$ with some additional gateway nodes and there exists
an injective mapping $m:V\rightarrow V_r$ such that, $\forall v\in V$, $Type(v)
= Type(m(v))$;

\item $\forall (u,v) \in L \ \exists g_u,g_v \in V_r$ such that $
Type(g_u) = Type(g_v) = gateway \wedge (g_u,m(u)) \in L_r \wedge (g_v,m(v))
\in L_r$ and there exists a path in $Tr$ between  $g_u$ and $g_v$ containing
only gateway nodes;
\item for each $v\in V$ if $\texttt{prox\_to\_src}(v) = 1$ then $Domain(m(v)) =
d_{src}$ and if $\texttt{prox\_to\_dst}(v) = 1$ then $Domain(m(v)) = d_{dst}$;
\item for each tuple $(d,t,m,n)$ represented by $C$ such that the type $t$
appears (as label of a node) in $T(V,L)$,  $m\leq \texttt{Num}(Tr,d,t) \leq n$ holds,
where $\texttt{Num}(Tr,d,t) = | \{ v | v\in Tr,\  Type(v) = t\hbox{ and }Domain(v) = d\}
|$.

\end{enumerate}

%
%
\end{definition} 

Note that, as indicated in item \emph{iv}), we assume that the domain
constraints refer to the VNF specified in the $\texttt{vnfList}$ provided by the
user.


We are now ready to state formally our configuration problem.

\begin{definition}[\probm]
\label{def:probm} Given a graph $G(V,L)$ that represents a network architecture,
an SFC user request $R$ and domain constraints $C$, the \probm consists in
finding an \solutn that satisfies the request $R$ and the constraint $C$. Such an
\solutn, if it exists, is called an admissible solution.  Furthermore, the
optimal \probm consist in finding an admissible solution $G(V',L')$which
minimize the following cost function: $\sum_{l \in  {L'}} c_l$. In this case the
solution found is called optimal \solutn.
\end{definition}

The following result shows that the problem that we are considering here is a difficult one.  
The proof is omitted for space reason and can be done by the reduction of 
the $k$-minimum spanning tree problem which is known to be  NP-hard
\cite{ravi1996spanning}.


\begin{theorem}[NP-hardness]
The optimal  \probm is NP-hard.
\end{theorem}

\section{SFC modeling with Constraint Programming}
\label{sec:problem_modeling}

%

In order to solve our \probm we translate it into a MiniZinc \cite{nethercote2007minizinc} finite domain specification. MiniZinc is a high level, solver independent, constraint modeling language which is widely used and is supported by large variety of constraint solvers. We assume some familiarity with MiniZinc and we invite the reader to consult \cite{nethercote2007minizinc} for further details.

Our translation is a direct encoding of the SFC design problem as defined in Section~\ref{sec:problem_def} in MiniZinc constraints. More precisely, we first model in terms of the MiniZinc language the network architecture and then we translate in MinZinc the user request and the domain  constraints defined in the JSON format. The MiniZinc specification of the network architecture  is a straightforward translation of the graph described in the previous section and is provided below (comments are  indicated by $\%$).


\begin{lstlisting}[escapeinside={<}{>}]
  int: n_nodes;     <\hid{\% Number of nodes (VNFs).}>
  int: n_domains;   <\hid{\% Number of domains.}>
  int: n_node_links;<\hid{\% Number of arcs (links between nodes).}>
  int: M;           <\hid{\% Upper bound for arc costs.}>
  <\hid{\% Array containing cost of arcs between pairs of gateway nodes.}>
  array[1..n_domains, 1..n_domains] of 0..M: 
  domain_link_costs; <\hid{\% Array representing the arcs.}>
  array[1..n_node_links, 1..2] of 1..n_nodes: 
  node_links; <\hid{\% Array describing the properties of the nodes, }>
  <\hid{\% i.e. node id, the type of node, its domain}>
  array[1..n_nodes, 1..3] of int: nodes;
\end{lstlisting}

%
Upon a user request expressed in the intent format, we use a script to extract necessary information and by using $\texttt{dupList}$ we parses the $\texttt{vnfList}$ into $\texttt{vnf\_arcs}$ that represents the arcs of $\texttt{request-tree}$ and finally we create an instance for MiniZinc.

As for the specification of the SFC request and domain constraints, described in definitions \ref{def:userreq} and \ref{definition:VNFcon}  in terms of JSON specifications, we use a script to extract necessary information and by using $\texttt{dupList}$ we parses the $\texttt{vnfList}$ into the $\texttt{vnf\_arcs}$ array below. Analogously we parse the domain constraints to build the $\texttt{domain\_constraint}$ array and we obtain the following MiniZinc code:
\begin{lstlisting}[escapeinside={<}{>}]
  int: start_domain;
  int: target_domain;
  int: n_types;      <\hid{\% Number of VNF types except Gateway}>
  int: vnflist_size; <\hid{\% The length of vnflist}>
  int: n_dcons;      <\hid{\% Number of domain constraint}>
  <\hid{\% The order of VNF in the service request.}>
  array[1..vnflist_size] of 0..n_types vnflist 
  <\hid{\% arcs of request-tree derived from vnflist}>
  array[1..vnflist_size-1, 1..2] of 
  0..vnflist_size: vnf_arcs;
  <\hid{\% VNF service in start domain}>
  array[1..vnflist_size] of 0..1: 
  proximity_to_source;   
  <\hid{\% VNF service in target domain}>
  array[1..vnflist_size] of 0..1: 
  proximity_to_destination;  <\hid{\% Domain constraints containing: domain id,vnf types, min, max.}>
  array[1..n_dcons, 1..4] of int: 
  domain_constraints;
\end{lstlisting}
To model our problem we then introduce two groups of MiniZinc variables, the first representing the selection of arcs, links, domains and domain connection, and the second to ensure that the selected nodes corresponding to the VNFs in $\texttt{vnfList}$ and their order is feasible.
%
%
%
%
%
%
%
%
Next we introduce the constraints which can be classified into three groups: the first one states the relations between variables (a.k.a channel constraints), the second guarantees that the variable values meet the request requirements and the last one ensure the tree properties of the solution. The key variable among all is the variable $\texttt{link\_selection}$, it is possible to build a relation with it to any other variables, e.g. to specify if a node or domain is selected it is enough to say whenever a link is selected then the related nodes and their domains are selected. The details of this formalization are omitted for space reasons and can be found in \cite{minizinc_imp}.

With these  constraints we are able to obtain an admissible \solutn. 
The optimal solution is the obtained by optimizing the sum of domain link costs of 
among all possible admissible solutions.


\section{Empirical Validations}
\label{sec:experiments}

We now describe the validation experiments which we have conducted in order to compare the performance of different state-of-the-art solvers and to assess the efficiency and scalability of our approach.

As for the experiment setup, we have generated the dataset representing the network in a random way. 
We assume $n$ nodes  and $m$ domains with $\frac{n}{m} > 2$. 
We select $m$ out of the $n$ 
nodes  and consider them as $\texttt{gateway}$
while for the remaining nodes we associate 
randomly to each of them a  VNF type from the set of types assumed in this paper
(see Section~\ref{sec:problem_def}). Next we defined the arcs according to the
definition in Section~\ref{sub:SFCdesign}
with costs in the range $[1,100]$.
Regarding the SFC user request, we created a dataset of possible requests
that may occur in practice, which are compliant with the assumptions we made in the paper and with the ETSI specifications \cite{isg2015gs}, 
from which we randomly choose specific instances. 
We consider the number of nodes and the number of domains
as features that characterize the specific instance dimension.
For each instance dimension we generate 10 scenarios 
and for each scenario we generate 10 requests which will be 
performed sequentially. We record the response time, that is the time needed to 
find optimal solution or  to discover that the instance
is unsatisfiable, with a cutoff time as 5 seconds for each run. The experiments were run 
on a Debian cluster with machines equipped with
Intel Core$i5$ 3.30GHz and 8 GB of RAM.

\begin{figure*}[t]
    \centering
    \begin{subfigure}[b]{0.49\textwidth}
        \includegraphics[width=\textwidth]{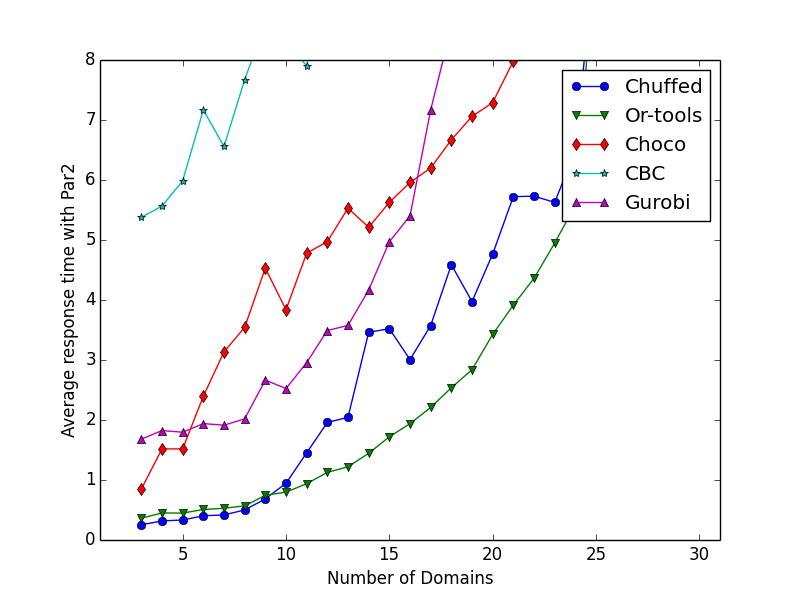}
        \caption{Response time with 300 nodes}
    \label{fig:varydomains_a}
    \end{subfigure}
    \begin{subfigure}[b]{0.49\textwidth}
        \includegraphics[width=\textwidth]{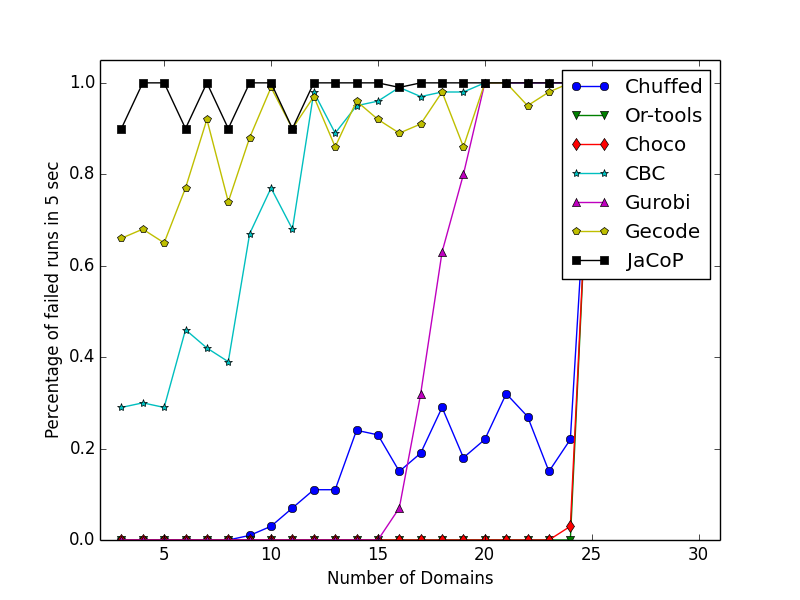}
        \caption{Percentage of Failed Runs}
        \label{fig:varingnodes}
    \end{subfigure}
    \caption{Solvers Comparison}\label{fig:comparison}
    \vspace{-0,5cm}
\end{figure*}

We first compared the performance of five different state-of-the-art CP
solvers, namely, Or-Tools $v6.7$~\cite{ortools2018}, Choco
$4.0.4$~\cite{choco}, JaCoP~\cite{kuchcinski2013jacop}
Gecode~\cite{schulte2006gecode}, Chuffed \cite{chu2010symmetries} and two
Mixed Integer Programming (MIP) solvers, Gurobi \cite{gurobi2015gurobi} (one
of the most performing MILP solvers \cite{gurobi75benchmark})  and
CBC~\cite{cbcref}\footnote{The Or-Tools were downloaded from Google OR
official page and other solvers were taken either from
SUNNY-CP~\cite{DBLP:conf/lion/AmadiniGM14,amadini2015sunny} or from the
MiniZinc distribution $v2.17$.} on the optimal \probm.
The solvers were run on scenario with 300 nodes and different number of
domains (from 3 to 30), each request was combined with 2 random domain
constraints. In the graph \ref{fig:comparison} (a) we show the response time
with $\texttt{Par2}$ penalty, where when a run was not completed at timeout we
consider its runtime as two times of the timeout (10 sec) \footnote{The
performance of Gecode and JaCoP were omitted since their performance were much
lower than those of the other solvers.}. Under the $\texttt{Par2}$ metric, it
can be seen that Chuffed and Or-Tools were the most competitive solvers in our
case, in particular, Chuffed runs faster with few number of domains (less than
10) while Or-Tools is more robust addressing instance with larger number of
domains. The part (b) of Fig \ref{fig:comparison} shows the percentage of runs
failed to prove optimality or unsatisfiability within timeout. It can be seen
that Choco and Or-Tools were the most competitive where they solved almost all
the instances with less than 24 domains. Chuffed started to have unsolved
instances when the number of domains goes beyond 9, however, it is still much
better that other solvers where they had failed runs even with 3 domains. It
worth noticing that the MIP solver Gurobi was less competitive than the CP
solver in our case, even though, the $\texttt{MIP/ILP}$ is the most popular
approach for NFV/SDN problems today.

\begin{figure*}[t]
    \centering
    \begin{subfigure}[b]{0.49\textwidth}
        \includegraphics[width=\textwidth]{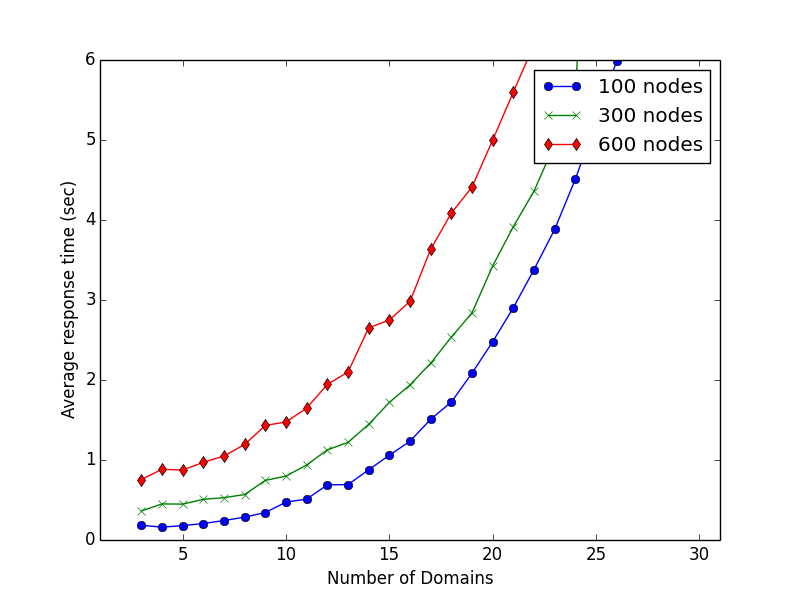}
        \caption{Response time in relation with number of domains}
    \label{fig:varydomains_b}
    \end{subfigure}
    \begin{subfigure}[b]{0.49\textwidth}
        \includegraphics[width=\textwidth]{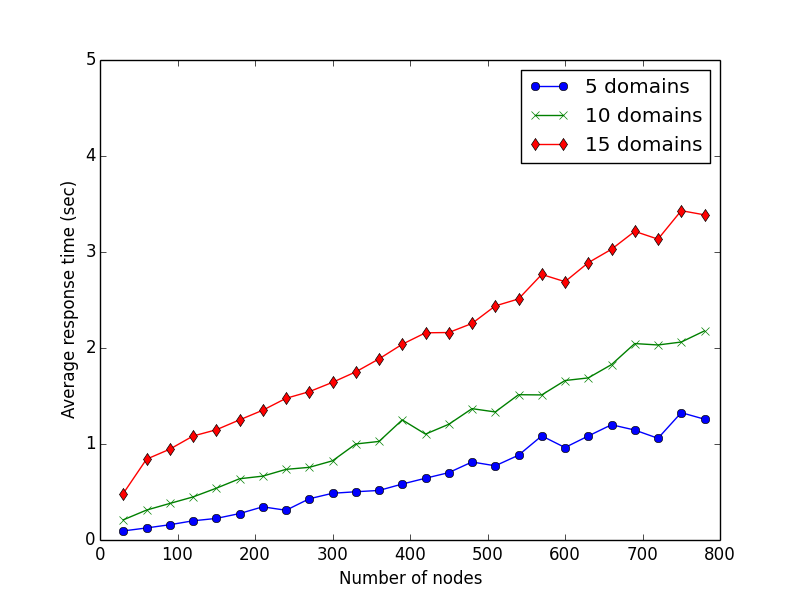}
        \caption{Response time in relation with number of nodes}
        \label{fig:varingnodes}
    \end{subfigure}
    \caption{System performance varying instance size}\label{fig:instancesize}
    \vspace{-0,5cm}
\end{figure*}

In the second set of experiments we considered only the solver Or-Tools and we
considered two groups of tests: fixing a number of nodes we vary the number of
domains from 3 to 30; fixing a number of domains we vary the number of nodes
from 30 to 800. In this case, the average runtime has excluded failed runs. As
one can see from Fig~\ref{fig:instancesize}, our application find the optimal
solution for  instances having more than 300 nodes and 10 domains in less then
a second\footnote{We also measured the runtime when request instance is
unsatisfiable, generally, it takes as much time as computing a satisfiable
instance.}.

Moreover, for the part (b) of the figure one sees that time grows 
almost linearly at the growth of node numbers. 
Since in practical applications one has hardly more than 10 domains and one has hardly a large number of 
nodes, and also, the links between domains are much less than our fully connected case, the results confirm that our system is relevant
to address the \probm and can scale up to consider large networks. It is worth mentioning that, for instance, the International Telecommunication Union in its Recommendation~\cite{ITU} sets an upper bound to the time needed to set up of a service at 7.5 seconds, well above the time needed here to solve the SF-Chaining problem. 


%

\section{Discussion and Conclusion}
\label{sec:related}


To the best of our knowledge the only other paper applying CP techniques to
programmable communication networks is~\cite{layeghy2016scor}, where the authors
consider the specific problem of optimizing the QoS of routing applications.
Here we consider a completely different problem, namely the definition of
expressive and efficient tools to solve the  Service Function Chaining design problem in general. There exists a large body of literature on the problem of mapping an SFC to the
(possibly virtualized) substrate network, optimizing some notion of
QoS. This problem, also called Service Function Chain Resource Allocation
(SFC-RA), has been mainly addressed with (Mixed) Integer Linear Programming
(M)ILP techniques. However, since in its full generality SFC-RA is an NP-hard
problem, many alternative approaches rely on approximated methods and
(meta)-heuristics (cf.~\cite{CommSurvSDN2014,CommSurvNFV2016,XieLWW16,HB16} for
more precise indications). 
When compared with other exact methods based on (M)ILP, CP provides a more
flexible and general approach. Since (M)ILP approaches consider a specific
formulation of the problem---customized for a narrow class of applications with
a specific function to be optimized---and require a large number of decision
variables and (in)equations, it becomes difficult to adapt existing solutions to
other cases.
Performance-wise, we cannot directly compare our work to other MILP based approaches, since
the problem we are solving here is more general than the specific ones treated
in the literature. However, our experimental results show that CP solvers are more efficient than MILP solvers on the problem we consider and support our claim that the
proposed model can scale efficiently. 

As future work, we will include our tool into a networking tool-chain for
directly applying synthesized SFC plans on target networks. We intend to
further investigate the definition of a high level, intent-based language for
SFC specification. Beside allowing to express quickly and intuitively SFC
requests, such an abstract language naturally would allow to use
modularization and typing~\cite{pierce2002types} principles with the following
benefits. First, support for the creation of libraries of standardized SFCs,
e.g., configurations that adhere to administrative regulations which can be
directly used with little customization effort. Second, the definition of
complex specifications obtained by combining simpler ones. Third, to
efficiently check if SFC specifications are well-formed (e.g., if the traffic
encrypted by a VPN is decrypted by a complementary function ) and if they
follow best practices (e.g., by warning users that, by using a VPN function
outside the domain of the source, the traffic might be exposed to attackers).

\bibliographystyle{ieeetr}
\bibliography{biblio}

\end{document}